\documentclass[a4paper,USenglish,cleveref, autoref, thm-restate]{lipics-v2021}

\usepackage{longtable}
\usepackage{nowidow}
\usepackage[ruled,lined,linesnumbered, noend]{algorithm2e}
\DontPrintSemicolon{}

\title{IBB:\@ Fast Burrows-Wheeler Transform Construction for Length-Diverse DNA Data.} 

\author{Enno Adler}{Paderborn University, Germany}{enno.adler@uni-paderborn.de}{https://orcid.org/0000-0001-8952-0325}{}

\author{Stefan Böttcher}{Paderborn University, Germany}{stefan.boettcher@uni-paderborn.de}{}{}

\author{Rita Hartel}{Paderborn University, Germany}{rita.hartel@uni-paderborn.de}{}{}

\author{Cederic Alexander Steininger}{Paderborn University, Germany}{cederics@mail.uni-paderborn.de}{}{}

\authorrunning{E. Adler et al.} 

\Copyright{Enno Adler, Stefan Böttcher, Rita Hartel and Cederic Alexander Steininger} 



\supplement{https://github.com/adlerenno/ibb}

\nolinenumbers 

\EventEditors{-}
\EventNoEds{1}
\EventLongTitle{arXiv}
\EventShortTitle{arXiv 2025}
\EventAcronym{arXiv}
\EventYear{2025}
\EventDate{February 03, 2025}
\EventLocation{Paderborn, Germany}
\EventLogo{}
\SeriesVolume{1}
\ArticleNo{1}

\newcommand{\bigO}{\mathcal{O}}

\begin{document}

\maketitle

\begin{abstract}
    The Burrows-Wheeler transform (BWT) is integral to the FM-index, which is used extensively in text compression, indexing, pattern search, and bioinformatic problems as \textit{de novo} assembly and read alignment. Thus, efficient construction of the BWT in terms of time and memory usage is key to these applications. We present a novel external algorithm called \textit{Improved-Bucket Burrows-Wheeler transform} (IBB) for constructing the BWT of DNA datasets with highly diverse sequence lengths. IBB uses a right-aligned approach to efficiently handle sequences of varying lengths, a tree-based data structure to manage relative insert positions and ranks, and fine buckets to reduce the necessary amount of input and output to external memory. Our experiments demonstrate that IBB is 10\% to 40\% faster than the best existing state-of-the-art BWT construction algorithms on most datasets while maintaining competitive memory consumption.
\end{abstract}

\section[Introduction]{Introduction}

The Burrows-Wheeler transform (BWT)~\cite{burrows1994} is a widely used reversible string transformation with applications in text compression, indexing, and pattern search~\cite{ferragina2005}. The BWT reduces the number of equal-symbol runs for data compressed with run-length encoding and allows pattern search in time proportional to the pattern length~\cite{ferragina2005}. Because of these advantages and the property that the BWT of a string $S$ can be constructed and reverted in $\bigO(|S|)$ time and space, the BWT plays an important role in computational biology, for example, in \textit{de novo} assembly~\cite{simpson2012, langmead2012} and short-read alignment. The tools BWA~\cite{Li2010}, Bowtie2~\cite{langmead2009, langmead2012}, MICA~\cite{luo2015}, and SOAP2~\cite{li2009} showcase the importance of the BWT.

Various suffix sorting algorithms were created to fit the diverse properties of strings or DNA sequences. For example, BCR~\cite{bauer2013} was designed for short-read BWT construction, whereas ropeBWT2~\cite{Li2014} focuses on long reads. See Puglisi et al.~\cite{puglisi2005} and Dhaliwal et al.~\cite{dhaliwal2012} for an overview of suffix sorting algorithms; in \autoref{section:related_work} we summarize the state-of-the-art of BWT construction for DNA sequences.

In this paper, we present our new BWT construction algorithm, \textit{Improved-Bucket Burrows-Wheeler transform} (IBB), which is especially designed for collections of strings $W_{i}$ of highly diverse length. In \autoref{tab:bwt_construction_algortihms}, some of the listed datasets have very diverse length distributions. 

Herein, our main contributions are as follows: 

\begin{itemize}
    \item An algorithm, IBB, to construct the BWT for DNA sequences of datasets with high diversity in length.\footnote{The implementation is available at \url{https://github.com/adlerenno/ibb}.}
    \item An extensive comparison of state-of-the-art BWT construction algorithms regarding the construction time and memory usage of BWT construction.\footnote{The test is available at \url{https://github.com/adlerenno/ibb-test}.} We show that IBB using a RAM disk is 10\% to 40\% faster than all other algorithms on most datasets while also using an amount of RAM smaller than most other approaches.
\end{itemize} 

\section[Related Work]{Related Work} \label{section:related_work}

The BWT~\cite{burrows1994} is key to many applications in bioinformatics such as short read alignment and \textit{de novo} assembly. Thus, there is a broad field of ideas how to compute the BWT for a collection of DNA strings. As the BWT can be computed by taking the characters from the suffix array at the position before the suffix, suffix array construction algorithms (SACAs) like divsufsort~\cite{fischer2017}, SA-IS by Nong et al.~\cite{nong2011}, gSACA-K by Louza et al.~\cite{Louza2016}, or gsufsort by Louza et al.~\cite{Louza2020} can compute the BWT.\@ Many SACAs rely on induced suffix sorting, which iteratively induces the order of suffixes from a sorted suffix subsets. Induced suffix sorting achieves its popularity by being a practically efficient linear-time suffix sorting algorithm.

The grlBWT method by D{\'{\i}}az{-}Dom{\'{\i}}nguez et al.~\cite{navarro2023} is an example for a BWT construction algorithm that uses induced suffix sorting. It also uses run-length encoding and grammar compression for intermediate results, which results in a faster BWT construction. 

The approach eGap by Egidi et al.~\cite{egidi2019} uses a divide and conquer strategy on the input collection: It divides the collection into subcollections and computes the BWT for the subcollection using gSACA-K~\cite{Louza2016}. After creating the BWT for each subcollection, eGap merges them to get the final BWT.\@

BigBWT by Boucher et al.~\cite{boucher2019} and r-pfbwt by Oliva et al.~\cite{oliva2023} use prefix-free parsing to reduce the size of the input. Prefix-free parsing represents the input by a parse and a dictionary, which is a prefix-free set. In prefix-free sets, no two words from the set are prefixes of each other. Therefore, the lexicographical order of words can be determined without relying on the word length and thus, the order of two suffixes starting with words from the prefix-free set is determined by these words. BigBWT uses prefix-free parsing on the input once. The approach r-pfbwt extends BigBWT by applying prefix-free parsing on the parse again, which recursively reduces the necessary space to represent the input.

All prior mentioned approaches use a concatenation of the input and a single-string construction in different ways. In contrast, using the \textit{Last-to-First-Mapping} (LF-Mapping), which is normally used to inverse the BWT, poses different approaches as well: The BCR algorithm by Bauer et al.~\cite{bauer2013} inserts one character of each sequence per iteration for a collection of short DNA reads. The LF-mapping together with current insert position and the partially constructed part of the BWT is used to compute the next insert position of the next symbol of a sequence.\@ BCR partitions these partially constructed BWTs into buckets, where one bucket contains all symbols prior to suffixes starting with the same symbol.\@ BCR is an external algorithm, as it stores the buckets in external memory. Ropebwt and ropebwt2 by Li~\cite{Li2014} resemble BCR, but they employ $B+$ trees for the buckets and are in-memory algorithms. Our approach, IBB, uses the LF-mapping as well, but introduces finer buckets and uses right-alignment on the input collection to reduce the size of partially constructed BWTs.



\section[Preliminaries]{Preliminaries}

In \autoref{appendix:abbreviations}, there is a list of symbols used in this paper.

We define a string $S$ of length $|S| = n$ over $\Sigma = \{A, C, G, T\}$ by $S = S[0] \cdot S[1] \cdots S[n-1]$ with $S[i] \in \Sigma$ for $i < n$. We write $S[i, j] = S[i] \cdot S[i+1] \cdots S[j]$ for a substring of $S$, $S[i,j] = \epsilon$ if $i > j$, and $S[i..] = S[i, |S|-1]$ for the suffix starting at position $i$. 
We define $rank_S(x, c) = |\{i < x : S[i] = c\}|$ as the number of times $c$ occurs in $S[0, x-1]$. Next, we define $count_S(c) = |\{0 \leq i < |S| : S[i] < c\}|$ to be the number of characters in $S$ that are lexicographically smaller than $c$. 

Given a collection $(S_i)_{0\leq i < m}$ of $m$ strings of any length over $\Sigma$, let $M = max(\{|S_i| : 0 \leq i < m\})$, and we set $W = S_0 \$_0 \dots S_{m-1} \$_{m-1}$ with $\$_0 < \dots < \$_{m-1} < A < C < G < T$. We call the elements $S_i$ of $(S_i)_{0\leq i < m}$ words. 

The Burrows–Wheeler transform $BWT(W)$~\cite{burrows1994} of $W$ can be computed
by taking the last column of the sorted rotations of $W$. We define the LF-Mapping as \[LF(i) = rank_{BWT(W)}(i, BWT(W)[i]) + count_{BWT(W)}(BWT(W)[i]).\] For $0 \leq j < m$, the string $S_j$ is obtainable from $BWT(W)$ by computing \[BWT(W)[LF^{|S_j|-1}(j)] \dots BWT(W)[LF(j)]\cdot BWT(W)[j] = S_j\] right to left and by aborting at $BWT(W)\left[LF^{|S_j|}(j)\right] = \$_{j-1}$ or $BWT(W)\left[LF^{|S_0|}(0)\right] = \$_{m-1}$ in case of $j = 0$ because the length $|S_j|$ is unknown. Since IBB does not need to differentiate the $\$_j$ symbols, we can simply write $\$$ for each of them.


\section[IBB]{IBB}

\subsection{Right Alignment}

We construct $BWT(W)$ without concatenating $W$ and without using a single-string BWT construction algorithm. Instead, like BCR and ropeBWT2, we use iterations $t = 0, \dots, M$. In each iteration, we insert at most one character of each word to compute a partial BWT string $bwt(t)$. Furthermore, like BCR and ropebwt2, we insert the characters of each word in reverse order. However, we do not start to insert the last character of each word $S_j$ in iteration $t=0$. Instead, we begin inserting characters of longer words in earlier iterations with the goal to terminate the insertion after the insertion of the first symbol of each word $S_j$ in the second last iteration $t=M-1$ followed by the insertion of a $\$$ character for each word $S_j$ in the last iteration $t=M$.

As we will process the strings $S_j$ from the end to the beginning, in each iteration $t$, we access the symbols $S_j[M-1-t]$ of strings $S_j$ with $|S_j|+t > M$. In the following, we write the shortcut $W_j[t]$ for $S_j[M-1-t]$. We also set $W_j[M] = \$$ for all words $W_j$. \autoref{figure:bwt-tree:words} shows the words $W_0, \dots, W_8$ already in reverse, right-aligned order during the iterations $t = 0, \dots, 10$.


Our insertion order has two advantages: First, we keep the partial $bwt(t)$ smaller compared to starting all words in iteration $t=0$, which allows faster computation of $bwt(t+1)$ from $bwt(t)$. Second, we only have to consider the 4 characters $A$, $C$, $G$, and $T$ instead of the 5 characters $A$, $C$, $G$, $T$, and $\$$ in $bwt(t)$ for all $t = 0, \dots, M-1$, which allows us to use smaller data structures.

The words $W_0, \dots, W_m$ are not necessarily ordered by their length and swapping their order would change $BWT(W)$. We start inserting a word $W_j$ in iteration $t = M - |W_j|$ and call the word \textit{active} then. Let the number of currently active words in iteration $t$ be $\alpha(t)$. When we start inserting a word $W_j$, we need to compute the insert position $P_j(M-|W_j|)$ for $W_j$ in iteration $t = M-|W_j|$, because the insertion of shorter words with an index $i < j$ in $(W_i)_{0\leq i < m}$ has not yet started. Our solution to this insert-position-problem uses a bit-vector $SB(t)$ of length $m$. Initially, $SB(0)[j]=0$ for all $0 \leq j < m$ and we set $SB(t)[j] = 1$ when $W_j$ starts in iteration $t = M - |W_j|$. After updating $SB(t)$ in an iteration $t = M - |W_j|$, the insert position $P_j(t)$ of $W_j[t]$ is $rank_{SB(t)}(j, 1)$, which counts the number of active words $W_z$ with index $z \in [0, j-1]$. For example, in iteration $t = 6$, the word $W_7$ is started. Its insert position shown in \autoref{figure:bwt-tree:active-words} is $P_7(6) = rank_{SB(6)}(7, 1) = 2$.

\subsection{Insert Position}

As in Ferragina et al.~\cite{ferragina2010}, the relative order of the symbols in $bwt(t)$, $t \geq 0$, is unchanged when inserting the new symbols to get $bwt(t+1)$. Thus, given the last insert position $P_j(t)$ of character $W_j[t]$, $P_j(t+1)$ can be obtained from an adjusted LF-Mapping. The adjustment regards the number of active words $\alpha(t+1)$ in iteration $t+1$ and is necessary, because $bwt(t)$, $t<M$, is not a valid BWT due to the missing $\$$ symbols. We need to adjust by $\alpha(t+1)$ instead of $\alpha(t)$ because within a non-partial BWT, every active word in iteration $t+1$ would need a $\$$ symbol.

The next insert position is $P_j(t+1) = LF(P_{j}(t)) + \alpha(t+1)$ for $W_j$ in iteration $t+1$ for $t < M$. In the last iteration $t = M$, we do not calculate any next insert positions, because we terminate after iteration $t = M$. Inserting the next character of each active word $W_j$ into $bwt(t)$—sorted according to the insert positions $P_{j}(t)$—yields $bwt(t+1)$. 

\begin{figure}
    \centering
    \begin{subfigure}{0.39\linewidth}
        \centering
        \includegraphics[width=\linewidth]{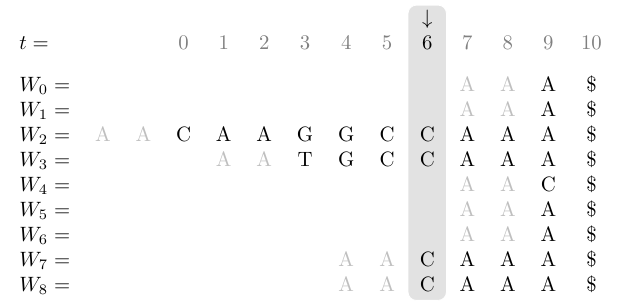}
        \caption{The Words $W_j$ are right aligned. The grey $A$s in front of a word $W_j$ are the predecessor sequence, which are introduced by $AW_j$.}\label{figure:bwt-tree:words}
    \end{subfigure}
    \hfill
    \begin{subfigure}{0.59\linewidth}
    \centering
        \includegraphics[width=\linewidth]{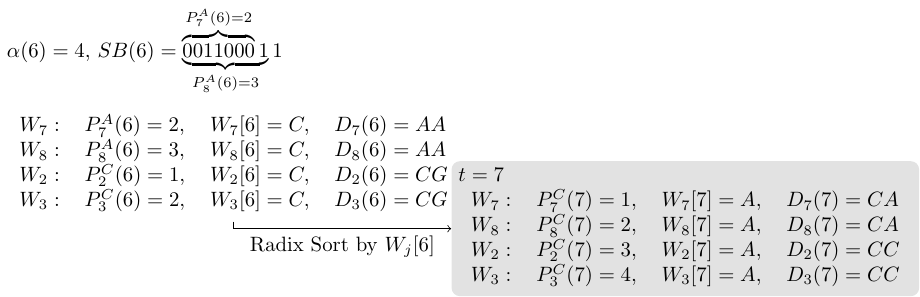}
        \caption{Active words of iteration $t = 6$ and $t = 7$, their insert positions $P$, their characters to insert, and their predecessor sequences $D$. The bitvector $SB$ is used to obtain the insert positions for the new started words $W_7$ and $W_8$.}\label{figure:bwt-tree:active-words}
    \end{subfigure}
    \hspace{0.1cm}

    \begin{subfigure}{\linewidth}
        \centering
        \includegraphics[width=\linewidth]{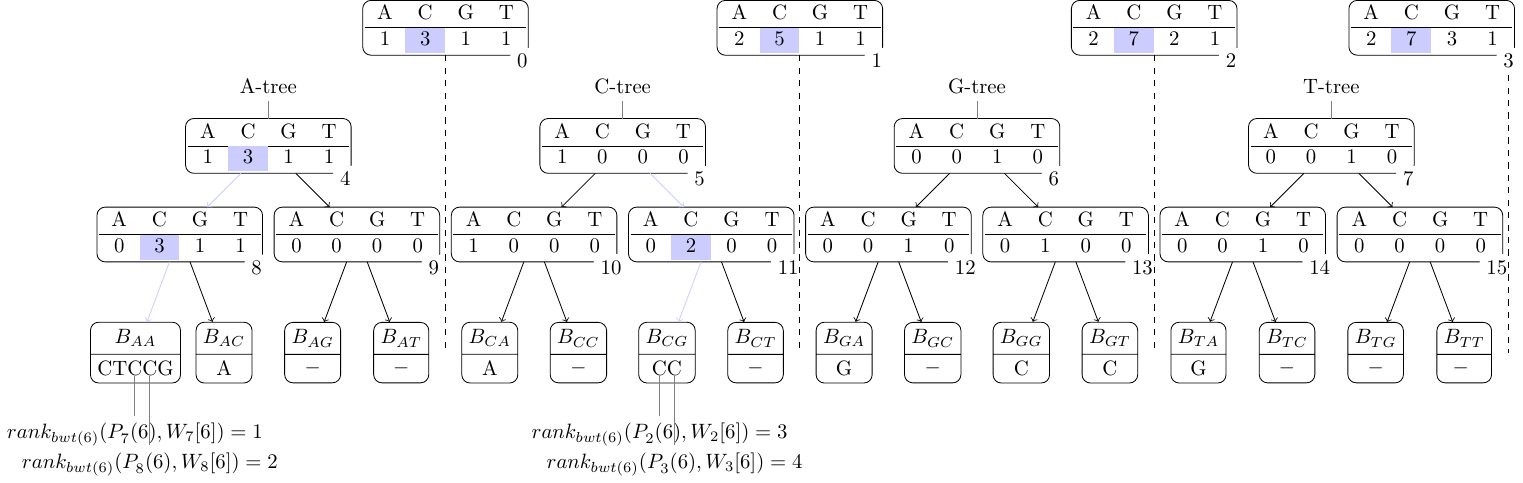}
        \caption{Tree structure for $k = 2$. Blue cells indicate an increase of the value in iteration $6$.}\label{figure:bwt-tree:tree-structure}
    \end{subfigure}
    \hspace{0.02cm}

    \begin{subfigure}{\linewidth}
        \centering
        \includegraphics[width=0.85\linewidth]{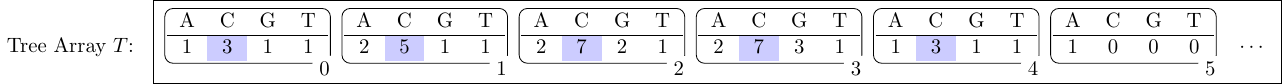}
        \caption{Tree array $T$. The numbers in the lower right corners of boxes are the indices within the $T$ array.}\label{figure:bwt-tree:tree-array}
    \end{subfigure}
    \caption{IBB structure for $k = 2$. All subfigures show the iteration $t = 6$ of the construction of the words in (a).}
    \label{figure:bwt-tree}
\end{figure}

In the example of \autoref{figure:bwt-tree:tree-structure}, the next insert position $P_3(7)$ of $W_3$ in iteration $t+1=7$ given $P_{3}(6) = 8$ and $bwt(6) = CTCCGAACCGCCG$ is 
\begin{align*}
    P_3(7) 
    &= LF(P_{3}(6)) + \alpha(7) \\
    &= rank_{C\textcolor{gray}{T}CC\textcolor{gray}{GAA}C\dots}(8, C) + count_{\textcolor{gray}{CTCCG}AA\textcolor{gray}{CCGCCG}}(C) + \alpha(7)\\ 
    &= 4 + 2 + 4 = 10.
\end{align*}

\subsection{Buckets}

BCR~\cite{bauer2013} and ropeBWT2~\cite{Li2014} segment $BWT(W)$ into buckets $B_c$ based on the number of a character $c$ in $W$. Formally, $B_c[i] = BWT(W)[i + count_W(c)]$ and $BWT(W) = B_\$ B_A B_C B_G B_T (B_N)$ in case of ropeBWT2s alphabet that includes $N$ additionally. We call the buckets $B_c$ \textit{level-1 buckets} and refine these buckets by looking at the first $k$ characters instead of one character $c$. Hereby, $k$ with $2 < k < log_4(|W|)$ is a fixed natural number. As shown in \autoref{figure:bwt-tree:tree-structure} for $k=2$, we have the \textit{level-k buckets} $B_{c_1 \cdots c_k}$ and we call $c_1 \cdots c_k$ the \textit{predecessor sequence} of the symbol to be inserted into the bucket. We omit special buckets for any predecessor sequence containing $\$$-symbols from $W$ and replace the character $\$$ and all following characters by $A$s for the definition of the buckets, so we use only characters $c_i \in \Sigma$ as predecessor symbols. The reason for this decision is that $\$$-symbols do not occur within a word $W_i$ and thus, some level-k buckets, for example $A\$C$ for $k=3$, would be empty by construction. 

Like BCR~\cite{bauer2013} and ropeBWT2~\cite{Li2014}, we can properly define the buckets of the final $BWT(W)$. For that purpose, we define $AW_j = (A^{k} \cdot W_j)$ as the word $W_j$ prepended by $k$ $A$ symbols before the word $W_j$. We transfer the index $i$ of each symbol in $W_j$ to $AW_j$, so $AW_j[i] = W_j[i]$ for every word $W_j$. In \autoref{figure:bwt-tree:words}, the grey $A$s before the words $W_j$ belong to $AW_j$. For each predecessor sequence $D \in \Sigma^k$, let $C_W(D) = \sum_{j=0}^{m-1} |\{i : M-|AW_j|-1 \leq i < M \wedge{} AW_j[i,i+k-1] < D\}|$ be the number of predecessor sequences that are lexicographically smaller than $D$, obtained from the $k$-length substrings of $AW_j$. Then, we get that $BWT(W) = B_{A^k} B_{A^{k-1}C} B_{A^{k-1}G} B_{A^{k-1}T} B_{A^{k-2}CA} \cdots B_{T^{k-1}G} B_{T^k}$ is a well-defined partition into $4^k$ segments where a bucket is defined as $B_{D}[i] = BWT(W)[i + C_W(D)]$.

To properly insert the character $W_j[t]$ into its corresponding bucket, we need to know the predecessor sequence $D_j(t)$ for $W_j[t]$ in iteration $t$. Because the BWT can be obtained by taking the last column of the sorted rotations and the buckets are ordered, the predecessor sequence $D_j(t)$ occurs at the start of the rotation of the character $W_j[t]$, if the $\$$ is replaced with $k$ $A$ symbols. Thus, $D_j(t) = AW_j[t-1] \cdots AW_j[t-k]$ is the proper predecessor sequence for character $W_j[t]$. 

Since $D_j(t) = AW_j[t-1] \cdots AW_j[t-k]$, $D_j(t)$ are the last $k$ inserted characters of word $W_j$ with the last inserted character $W_j[t-1]$ first or, if $t<k$, $D_j(t)$ is prepended by $k-t$ $A$s. Especially, you can get $D_j(t+1) = W_j[t] \cdot D_j(t)[0,k-2]$ by shifting the previous predecessor sequence $D_j(t)$ and placing the last inserted character $W_j[t]$ at index $0$. For example, in \autoref{figure:bwt-tree:active-words}, $D_2(7) = CC$, because we insert $W_2[7-1] = C$ in the $CG = D_2(6)$ bucket, so $D_2(7) = W_2[6]\cdot D_2(6)[0,k-2] = C \cdot (CG)[0,0] = CC$.

The idea is that knowing the bucket $B_{W_j[t-1]}$ to insert into and the insert position $P_{j}^{W_j[t-1]}(t)$ within that bucket is sufficient for insertion and that the global insert position $P_{j}(t)$ is not needed for insertion. To calculate the next insert position $P_j(t+1)$, we need $rank_{bwt(t)}(P_{j}(t), W_j[t])$ and $count_{bwt(t)}(W_j[t])$ for the LF-mapping. Using buckets and predecessor sequences, we can omit computing (or maintaining) and adding $count_{bwt(t)}(W_j[t])$: Let $P_{j}^{W_j[t-1]}(t)$ be the insert position of character $W_j[t]$ where we insert the character $W_j[t]$ into the bucket $B_{W_j[t-1]}$. Bauer et al.~\cite{bauer2013} prove $P_{j}^{W_j[t-1]}(t) = rank_{bwt(t-1)}(P_{j}(t-1), W_j[t-1])$ using a different notation. In case of $W_j[t] = A$, we get the next insert position $P_{j}^{A}(t+1) = rank_{bwt(t)}(P_{j}(t), A) + \alpha(t+1)$ due to our merge of the $\$$ bucket into the $A$ bucket. In other words, $count_{bwt(t)}(W_j[t])$ is equal to the total number of characters in the buckets of smaller characters $c < W_j[t]$. If $W_j[t] = A$, $count_{bwt(t)}(A)$ is the number of active words $\alpha(t+1)$, which would be the size of the bucket $B_{\$}$ in $bwt(t+1)$ without the merge of the bucket $B_{\$}$ into the bucket $B_A$.

\subsection{Trees} \label{subsection:trees}

In order to save computation time for rank operations on previous buckets, we introduce four balanced binary trees, \textit{A-tree}, \textit{C-tree}, \textit{G-tree}, and \textit{T-tree}, of depth $2(k-1)$, as shown in \autoref{figure:bwt-tree:tree-structure} for $k = 2$. We save the balanced binary trees in an array $T[0, 2^{2k}-1]$ of length $4^{k} = \bigO(|W|)$ using the Eytzinger Layout: The root of the balanced binary trees, A-tree to T-tree, are saved in $T[4]$ to $T[7]$; the left child of a node $T[i]$ is stored at index $2i$ and the right child position at index $2i+1$ in $T$. Each internal node has $|\Sigma|=4$ counters, where the counter $T[i].c$ associated with $c\in\Sigma$ represents the number of characters $c$ in the left subtree of node $i$. The leaf nodes are the buckets, which we discuss in \autoref{subsection:leaves}.

The trees $T[4]$, $T[5]$, $T[6]$, and $T[7]$ represent the level-1 buckets $B_A$, $B_C$, $B_G$, and $B_T$, respectively. We continue to refer to each of these buckets as the concatenation of all level-k buckets of the corresponding tree.

We speed up the computation of $rank_{bwt(t)}(P_{j}(t), W_j[t])$ by storing additional information in $T[0]$ to $T[3]$. In $T[0]$, we store the total number of characters in the leaves of the A-tree. In $T[1]$, we store the number of characters of the C-tree plus the A-tree, so we have an accumulate count of the number of characters of the whole part of $bwt(t)$ prior to the start of the G-tree. In the same way, $T[2]$ and $T[3]$ represent the total number of characters up to the end of the G-tree and T-tree.
\autoref{figure:bwt-tree:tree-structure} visualizes the tree structure $T$. 

We use $T[0]$ to $T[3]$ to determine the number of characters $W_j[t]$ up to the beginning of tree corresponding to $W_j[t-1]$. Thereby, $rank_{bwt(t)}(P_{j}(t), W_j[t])$ can be determined by processing only the tree corresponding to $W_j[t-1]$. We define the \textit{level-1 rank} as $rank1(j, t) = rank_{B_{W_j[t-1]}}\left(P_{j}^{W_j[t-1]}(t), W_j[t]\right)$, which is the rank of symbol $W_j[t]$ inserted at its insert position $P_{j}^{W_j[t-1]}(t)$ on the tree of symbol $W_j[t-1]$.

\begin{align*}
    rank_{bwt(t)}(P_{j}(t), W_j[t])= \begin{cases}
        rank1(j, t) & D_j(t)[0] = A\\
        T[0].(W_j[t]) + rank1(j, t) & D_j(t)[0] = C\\
        T[1].(W_j[t]) + rank1(j, t) & D_j(t)[0] = G\\
        T[2].(W_j[t]) + rank1(j, t) & D_j(t)[0] = T\\
    \end{cases}
\end{align*}

\subsection{Navigation}

To get the local insertion position $P_{j}^{D_j(t)}(t)$ within level-k bucket $B_{D_j(t)}$, we navigate through the balanced binary trees in $T$ using the predecessor sequence $D_j(t) = c_1 c_2 \dots c_k$: While the predecessor symbol $c_1$ of each word determines the selection of one of the 4 binary trees, the predecessor symbols $c_2,\dots,c_k$ of $W_j[t]$ determine the navigation direction inside the $c_1$-tree towards the bucket $B_{D_j(t)}$. Explicitly, for the navigation within the first two levels of the $c_1$-tree, the symbol $c_2$ is used, for the next two levels $c_3$, and so on. We navigate twice left if $c_i = A$; first left and second right if $c_i = C$; first right and second left if $c_i = G$; and twice right if $c_i = T$. If a left navigation step at node $i$ is issued for $c = W_j[t]$, we need to increase $T[i].c$ by one, because we will insert the character $c$ into a leaf in left subtree of node $i$. 

If we want to insert symbols from different words at the same time and some symbols have to be inserted into a bucket left of an inner node $i$ while other symbols have to be inserted into a bucket right of the inner node $i$, the left steps must increase $T[i].c$ at a node $i$ before we can proceed with the right steps at node $i$. Otherwise, $T[i].c$ would not be the correct number of occurrences of the character $c$ in the left subtree of node $i$.

\begin{figure}
    \centering
    \includegraphics[width=\textwidth]{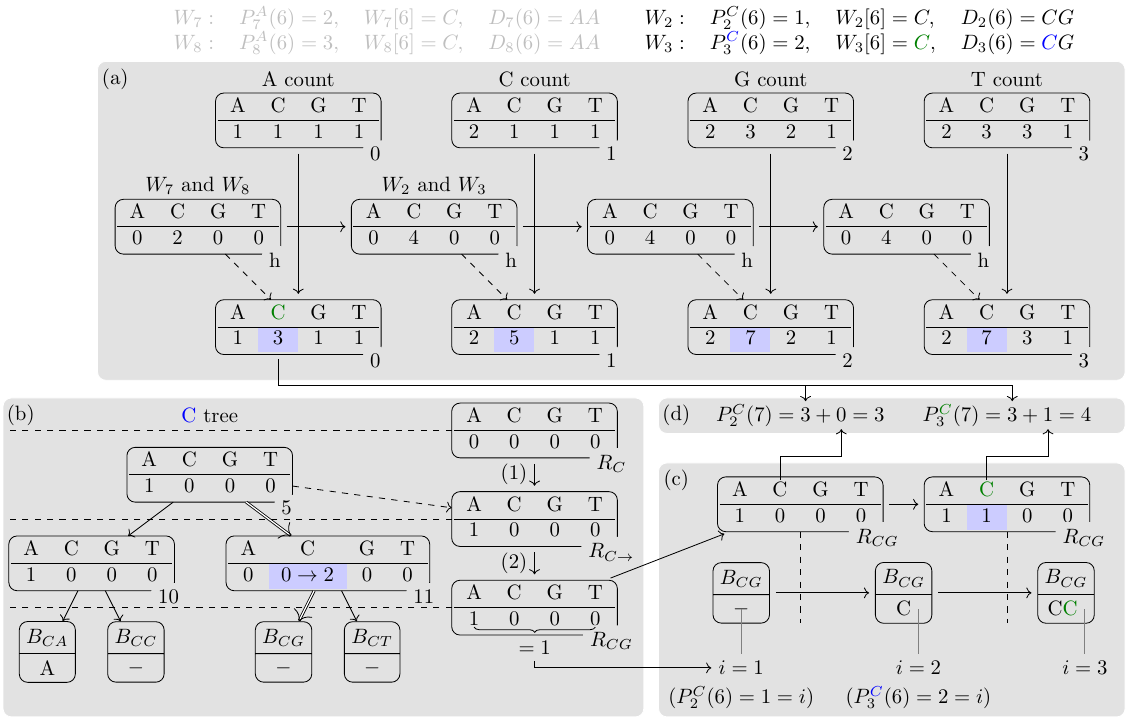}
    \caption{Insertion of $W_2[6]$ and $W_3[6]$ into IBB structure. Blue backgrounds mean that the numbers are changed. The green and blue $C$ highlight which $C$ is used in which location to determine the used tree or value.}
    \label{figure:bwt-tree-insert}
\end{figure}

For each right step, our goals are to transform the insert position $P_{j}^{W_j[t-1]}(t)$ into the local insert position $P_{j}^D(t)$, and to determine $rank1(j, t)$. 
The number of characters $c$ in the local buckets before the bucket $B_D$ is equal for all words of which characters are inserted into the bucket $B_D$. This equality is true both in the case of every single $c\in\Sigma$ and for the sum of all characters $c\in\Sigma$. We use accumulators $R_D.c$ to reduce the number of necessary additions in the right step: For each sequence $D$, $R_D.c$ is the sum of all counters $T[i].c$ where $i$ is a node with a right step issued by $D$. Hereby, $D$ must not be of length $k$ during computation and can contain a $\leftarrow$ or $\rightarrow$ symbol at the end to indicate the step of the last half-processed predecessor symbol. We only compute $R_D$ for predecessor sequences $D$ where at least one character $W_j[t]$ has $D$ as predecessor sequence in iteration $t$. Thus, we have at most $4 \cdot \min(4^k, m)$ different $R_D.c$ accumulators instead of exactly $4m$ if we would use four variables for each individual word. 

Using the accumulator $R_D$ for the character $W_j[t]$, so $D = D_j(t)$, the local insert position of $W_j[t]$ is $P_{j}^{D_j(t)}(t) = P_{j}^{W_j[t-1]}(t) - \sum_{c\in\Sigma} R_D.c$. Because $\sum_{c\in\Sigma} R_D.c$ is again constant for all words that insert into the bucket $B_D$, the sum $\sum_{c\in\Sigma} R_D.c$ is only computed once per predecessor sequence $D$. Furthermore, we do not subtract the sum $\sum_{c\in\Sigma} R_D.c$ from each $P_{j}(t)$. Instead, we initialize the position counter with $\sum_{c\in\Sigma} R_D.c$. An example for the insertion using $R_D.c$ is shown in \autoref{figure:bwt-tree-insert}c.

To compute $rank1(j, t)$ for the insert position $P_j(t+1)$, we use $R_D.c$ with $c = W_j[t]$, which is the number of symbols $c$ up to the bucket $B_D$. Then, we compute the \textit{level-k rank} $rank\text{k}(j, t) = rank_{B_{D_j(t)}}\left(P_{j}^{{D_j(t)}}(t), W_j[t]\right)$ during insertion into the bucket $B_{D_j(t)}$. Thus, we get the overall expression of the insert position $P_j^{W_j[t]}(t+1)$ for iteration $t+1$ from the iteration $t$ as:

\begin{align*}
    P_j^{W_j[t]}(t+1) = \begin{cases}
        \alpha(t+1) + R_{D_j(t)}.(W_j[t]) + rank\text{k}(j, t) & D_j(t)[0] = A\\
        T[0].(W_j[t]) + R_{D_j(t)}.(W_j[t]) + rank\text{k}(j, t) & D_j(t)[0] = C\\
        T[1].(W_j[t]) + R_{D_j(t)}.(W_j[t]) + rank\text{k}(j, t) & D_j(t)[0] = G\\
        T[2].(W_j[t]) + R_{D_j(t)}.(W_j[t]) + rank\text{k}(j, t) & D_j(t)[0] = T\\
    \end{cases}
\end{align*}

We parallelize the processing of the children each time both children need to be processed, as the further steps of the left and right child do not interfere with each other. We limit the number of threads close to the number of available processors on the used system, as this limit is most efficient.

\subsection{Sort Words}

If symbols from multiple words are inserted into the same bucket $B_D$, we must insert them in correct relative order. We insert the symbols from left to right; for that purpose, we sort the words $W_j$ from which the characters $W_j[t]$ are taken according to their predecessor sequence $D_j(t)$ and then according to their local insert position $P_{j}^{D_j(t)}(t)$ in bucket $B_{D_j(t)}$. The resulting sort order is equal if we sort according to the first predecessor symbol $D_j(t)[0] = W_j[t-1]$ first and then to the insert position $P_{j}^{W_j(t-1)}(t)$. We will use the latter ones.

In iteration $0$, all new started words are sorted by their insert position $P_j(0)$ by construction; thus they are also sorted according to their local insert position $P_{j}^{A^k}(0)$ because $P_j(0) = P_j^{A}(0) = P_j^{A^k}(0)$. Next, at the start of iteration $t$, we prepend all new active words $W_z$ to the sorted list of active words, because their predecessor sequence $D$ are $A^k$ and their insert positions $P_{z}^{A}(t)$ are lower than the insert positions $P_{j}^{A}(t)$ of the older active words $W_j$ due to the addition of $\alpha(t+1)$.

For iteration $t+1 > 0$, we sort the active words from iteration $t$ at the end of iteration $t$ by a single stable radix-sort step on $W_j[t]$. In \autoref{figure:bwt-tree:active-words}, the radix sort sorts the words according to the characters $W_j[6]$ resulting in the same order of active words in iteration $t=7$. The next insert positions $P_{j}(t+1)$ for one character $W_j[t+1]$ depends only on $rank_{bwt(t)}(P_{j}(t), W_j[t])$ for a group all words $W_j$ with the same character $c$ at position $t$, or on the constant term $\alpha(t+1)$ in case of $W_j[t] = A$. As the $rank$ is a monotonic non-decreasing function of the position, the sort-order of the group of all $W_j$ with $W_j[t] = c$ is stable. Thus, a single radix-sort step on $W_j[t]$ is sufficient to achieve the correct order of the active words $W_j$ of the iteration $t$.

\begin{table}
    \centering  
    \caption{Used BWT construction algorithms and datasets from \href{https://www.ncbi.nlm.nih.gov/}{NCBI}, \href{https://gage.cbcb.umd.edu}{GAGE}, or \href{https://downloads.pacbcloud.com/public/}{PacBio}. We used partDNA~\cite{adler2024} with parameter $h=4$ to partition the datasets TAIR10, GRCh38, GRCm39, and JAGHKL01 into sets of words. $avg(|W_i|)$ is the average length of words $W_i$; $M$ is the length of the longest word in $(W_i)_{0\leq i < m}$. We visualize the distribution of word lengths of GRCh38 in \autoref{figure:length-diversity}.}\label{tab:bwt_construction_algortihms}
    \begin{subtable}{0.3\linewidth}
    \centering
    \begin{tabular}{ll}
        approach & paper\\
        \hline

        \href{https://github.com/adlerenno/ibb}{IBB} & this paper\\
        
        \href{https://github.com/giovannarosone/BCR_LCP_GSA}{BCR} &\cite{bauer2013} \\
        
        \href{https://github.com/lh3/ropebwt}{ropebwt} & – \\

        \href{https://github.com/lh3/ropebwt2}{ropebwt2} & \cite{Li2014}\\

        \href{https://github.com/lh3/ropebwt3}{ropebwt3} & – \\

        \href{https://gitlab.com/manzai/Big-BWT}{BigBWT} & \cite{boucher2019} 
        \\

        \href{https://github.com/marco-oliva/r-pfbwt}{r-pfbwt} & \cite{oliva2023} \\

        \href{https://github.com/y-256/libdivsufsort}{divsufsort} & \cite{fischer2017} \\

        \href{https://github.com/ddiazdom/grlBWT}{grlBWT} & \cite{navarro2023} \\

        \href{https://github.com/felipelouza/egap}{eGap} & \cite{egidi2019} \\

        \href{https://github.com/felipelouza/gsufsort}{gsufsort} & \cite{Louza2020} \\
    \end{tabular}
    \end{subtable}
    \hfill
    \begin{subtable}{0.69\linewidth}
        \centering
    \begin{tabular}{lrrr}
        dataset & $avg (|W_i|)$ & $M$ & $m$ \\ \hline

        \href{https://www.ncbi.nlm.nih.gov/assembly/GCF_000001735.4/}{TAIR10} & 59 & 4,558 & 2,031,079 \\
        \href{https://www.ncbi.nlm.nih.gov/datasets/genome/GCF_000001635.27/}
        {GRCm39} & 83 & 31,851 & 31,890,467 \\
        \href{https://www.ncbi.nlm.nih.gov/datasets/genome/GCF_000001405.40/}{GRCh38} & 67 & 36,931 & 46,529,667 \\
        \href{https://www.ncbi.nlm.nih.gov/datasets/genome/GCF_018294505.1/}
        {JAGHKL01} & 121 & 65,807 & 118,749,573 \\
        \hline
        \href{https://www.ncbi.nlm.nih.gov/sra/SRR11092057}{SRR11092057} & 137 & 151 & 5,314,809 \\
        \href{https://ftp.ncbi.nih.gov/genomes/INFLUENZA/}{influenza} & 1590 & 2,867 & 817,587 \\
        \href{https://gage.cbcb.umd.edu/data/Hg_chr14/Data.original/}{HGChr14} & 99 & 101 & 18,610,954 \\
        \href{https://downloads.pacbcloud.com/public/dataset/Onso/Zymo_wastewater/fastqs/}{zymowastewater} & 143 & 150 & 73,996,930 \\

    \end{tabular}
    \end{subtable}
    
\end{table}

\subsection{External Storage of Level-k Buckets} \label{subsection:leaves}

We store each level-k bucket $B_D$ on external memory and insert into the bucket $B_D$ by alternating two files. We denote the two files by $F_{0}(B_D)$ and $F_{1}(B_D)$. Let $L_{D}(t) \in \{0, 1\}$ be the index of the file $F_{L_{D}(t)}(B_D)$ that we suppose to write to. Initially, we set $L_{D}(0) = 0$ for all $D$. If no word $W_j$ has $D$ as predecessor sequence, we skip the bucket $B_D$. Otherwise, we input $F_{1-L_D(t)}(B_D)$ as a stream. We output the symbols of the input stream to $F_{L_D(t)}(B_D)$ until we reach the next lowest insert position $P_{j}^{D}(t)$ of a word $W_j$. We then output $W_j[t]$ before continuing with the input stream and the next lowest insert position $P_{z}^{D}(t)$ of some word $W_z$. If no such word $W_z$ with $D$ as predecessor sequence exists, we output the rest of the input stream. Both, the input and output stream, are buffered for performance reasons. 
At the end, we flip $L_{D}(t)$ for the next iteration $t+1$ if and only if bucket $B_D$ was changed.

To get $rank\text{k}(j, t)$ for the next insert position $P_j^{W_j[t]}(t+1)$, we count the number of characters $W_j[t]$ that are written to file $F_{L_D(t)}(B_D)$ until the current local insert position $P_j^D(t)$. As $rank\text{k}(j, t)$ is added to the value of $R_D.c$ with $c=W_j[t]$, we can directly increment the counter $R_D.c$ during writing $c$ to the output file $F_{L_D(t)}(B_D)$. Thereby, we get $rank1(j, t)$ from $R_D.c$ with $c = W_j[t]$ just before outputting the symbol $W_j[t]$ at position $P_{j}^{D}(t)$.

In \autoref{figure:bwt-tree-insert}, the characters $W_2[6]$ and $W_3[6]$ are inserted into $bwt(5)$ in iteration $6$. First, in \autoref{figure:bwt-tree-insert}a, we update $T[0]$ to $T[3]$ by increasing $T[0].C$ by $2$ and $T[1].C$, $T[2].C$, and $T[3].C$ by $4$ by using an accumulator $h$ with four counters $h.c$. Second, in \autoref{figure:bwt-tree-insert}b, we process the C-tree, because the first predecessor symbol is $D_2(6)[0] = D_3(6)[0] = C$. We do a right and then a left child step because the second predecessor symbol is $D_2(6)[1] = D_3(6)[1] = G$. In \autoref{figure:bwt-tree-insert}c, the right step is marked with (1) and the left step is marked with (2). We initialize the accumulators $R_C.c$ with $0$ for all $c$. Because there is no word that requires a left child step at $T[5]$, we omit cloning $R_C$ into $R_{C\leftarrow}$ and just keep $R_C$ as $R_{C\rightarrow}$. Additionally, we add the counters of the root of the C tree at $T[5]$ to $R_{C\rightarrow}$ to $R_C$, so we properly have the number of characters in the left subtree of $T[5]$ in $R_C$. Then, we process the right child $T[2*5+1] = T[11]$. First, we increase for $W_2$ and $W_3$ $T[11].C = 0$ twice by 1 to $2$, because both are left steps. As no word requires a right step at $T[11]$, we can omit that branch. By navigating to the child with index $2*11=22$, we reach a leaf of the tree, thus we insert the characters $W_2[6]$ and $W_3[6]$ to that bucket $B_{CG}$. 

In \autoref{figure:bwt-tree-insert}c, we perform the insertion into the bucket $CG$. The position counter $i$ is initialized with value $1$, which is the sum of the accumulators $R_C.c$. Because $P_j^C(6) = 1 = i$, we insert the character $W_2[6] = C$ at the beginning of bucket $B_{CG}$. We increase $R_{CG}.C$ by $1$ for the inserted $C$ of word $W_2$. In \autoref{figure:bwt-tree-insert}d, we compute the next insert position $P_2^C(7)$ for $W_2$ before incrementing $R_{CG}.C$. The next insert position $P_2^C(7) = 3$ for $W_2$ is the sum of the number of $C$ before the C-tree $T[5]$, which is $T[0].C = 3$, and the number of $C$ within the $C$ tree up to that position, which is $R_{CG}.C = 0$. In the same way we insert $W_3[6]$ and get the next insert position $P_3^C(7) = 4$ for $W_3$ in iteration $t=7$. 

\section{Experimental Results}

\begin{figure}
    \centering
    \begin{subfigure}{0.54\textwidth}
        \centering
    \includegraphics[width=\textwidth]{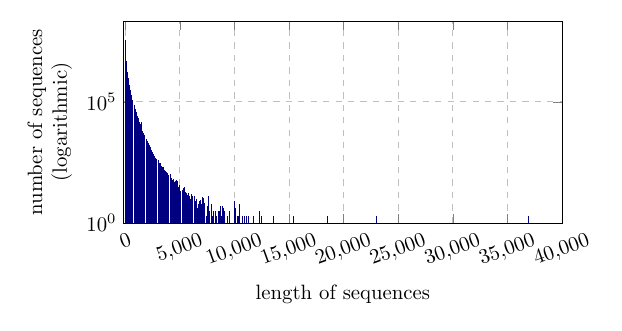}
    \caption{Distribution of the sequence length of the partitioned GRCh38 file. We summarize sequences over an interval of one-hundred characters. $91.26\%$ of the $46,529,667$ sequences are shorter than 200 characters, but there are two sequences between $36,900$ and $37,000$ characters.}
    \label{figure:length-diversity}
    \end{subfigure}
    \hfill
    \begin{subfigure}{0.435\textwidth}
    \centering
    \includegraphics[width=\textwidth]{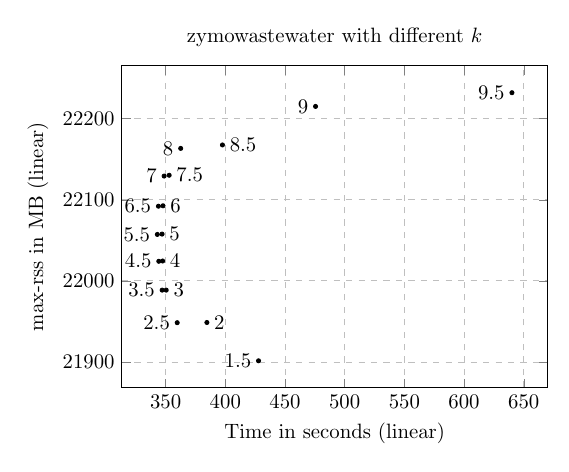}
    \caption{Construction time and RAM consumption of IBB for zymowastewater dataset tested for $1.5 \leq k \leq 9.5$. }
    \label{fig:k_results}
    \end{subfigure}
    \caption{}
\end{figure}

\begin{figure}
    \newcommand{\subfigurewidth}{0.455\textwidth}
    \centering
    \caption{BWT construction times and maximum resident set sizes (max-rss). A missing point means that the construction algorithm aborts or does not create an output file. }
    \label{fig:time_and_space}
    \begin{subfigure}[b]{\subfigurewidth}
        \centering
        \includegraphics[width=\textwidth]{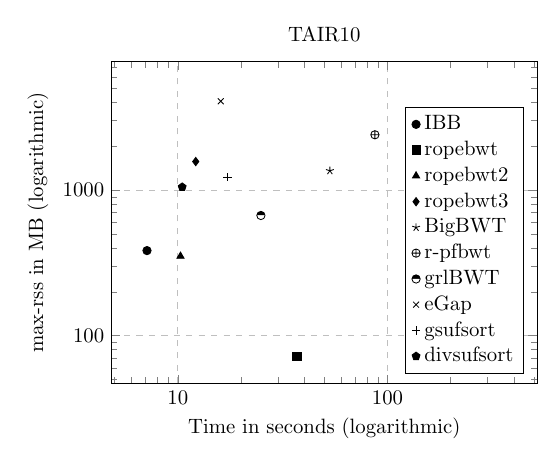}
    \end{subfigure}
    \hfill
    \begin{subfigure}[b]{\subfigurewidth}
        \centering
        \includegraphics[width=\textwidth]{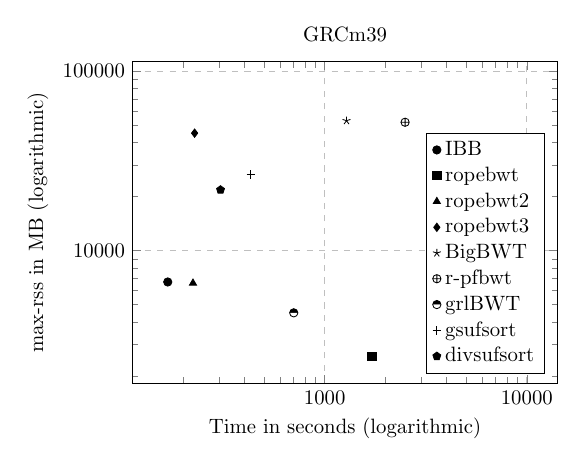}
    \end{subfigure}

    \begin{subfigure}[b]{\subfigurewidth}
        \centering
        \includegraphics[width=\textwidth]{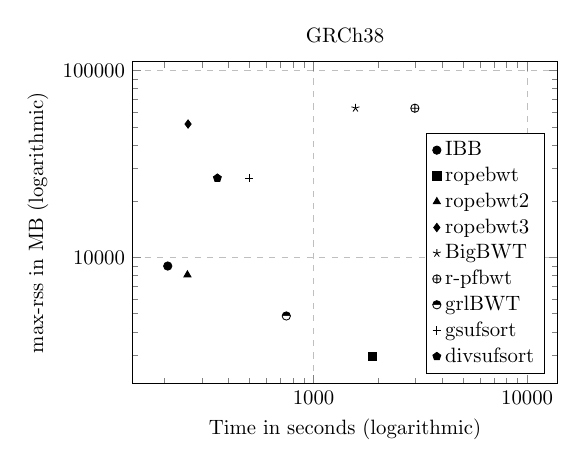}
    \end{subfigure}
    \hfill
    \begin{subfigure}[b]{\subfigurewidth}
        \centering
        \includegraphics[width=\textwidth]{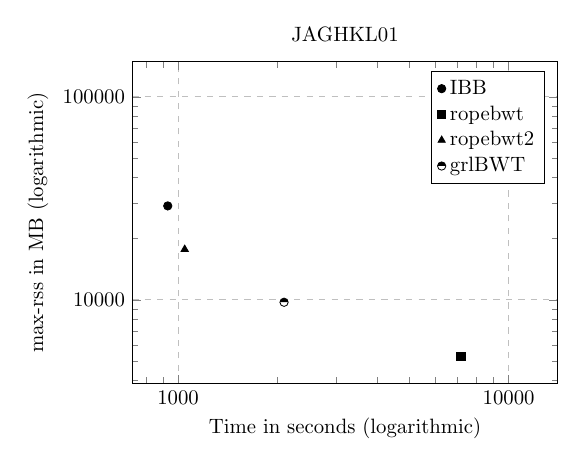}
    \end{subfigure}

    \begin{subfigure}[b]{\subfigurewidth}
        \centering
        \includegraphics[width=\textwidth]{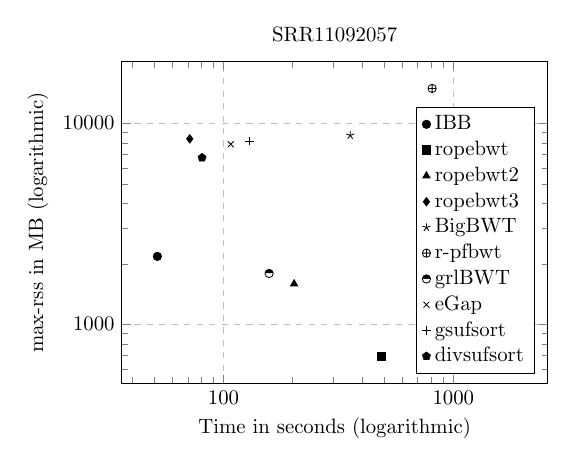}
    \end{subfigure}
    \hfill
    \begin{subfigure}[b]{\subfigurewidth}
        \centering
        \includegraphics[width=\textwidth]{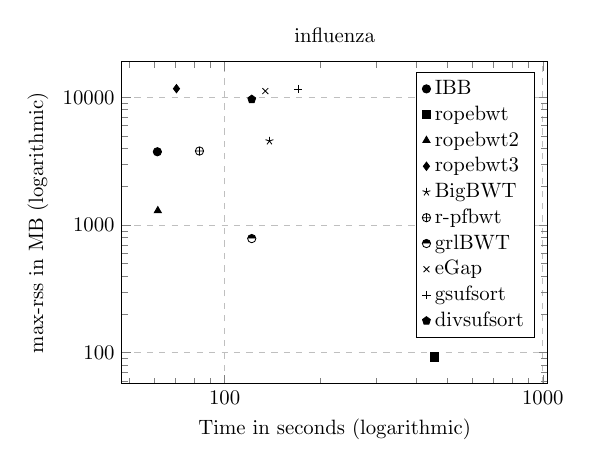}
    \end{subfigure}

    \begin{subfigure}[b]{\subfigurewidth}
        \centering
        \includegraphics[width=\textwidth]{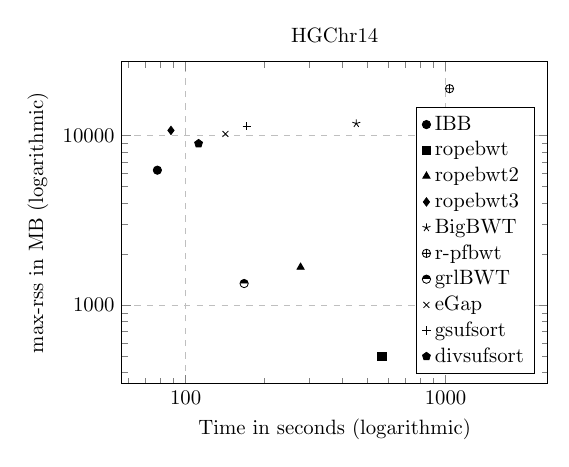}
    \end{subfigure}
    \hfill
    \begin{subfigure}[b]{\subfigurewidth}
        \centering
        \includegraphics[width=\textwidth]{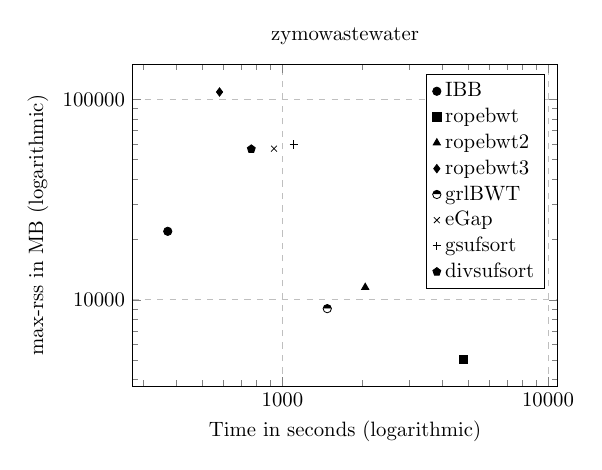}
    \end{subfigure}
\end{figure}

We compare the BWT construction algorithms on the datasets listed in \autoref{tab:bwt_construction_algortihms} regarding construction time and RAM usage. We obtain time, maximum resident set size (max-rss), amount of input (io\_in) and amount of output (io\_out) from Snakemake~\cite{moelder2021}.\footnote{The test is available at \url{https://github.com/adlerenno/ibb-test}.}  To obtain comparative results regarding construction time to the in-memory algorithms, we run all external-memory algorithms on RAM disks. Ambiguous bases in the datasets are omitted.
We performed all tests on a Debian 5.10.209-2 machine with 128GB RAM and 32 Cores Intel(R) Xeon(R) Platinum 8462Y+ @ 2.80GHz. 

We pass the number of 32 threads explicitly to IBB, grlBWT, r-pfbwt, BigBWT, and ropebwt3; other approaches, like grlBWT, that parallelize their construction, determine the number of threads on their own, because they have no such command line option. We pass the option \texttt{-R} to ropebwt, ropebwt2 and ropebwt3 to avoid processing the read sequences in both directions. We used the parameter options \texttt{w1 = 10}, \texttt{p1 = 100}, \texttt{w2 = 5}, and \texttt{p2 = 11} for r-pfbwt as suggested. Also, we did not force eGap into external, semi-external or internal memory mode.

BCR as an external algorithm designed for equal length datasets supports datasets of unequal length, nevertheless BCR failed on all datasets. GrlBWT failed using the RAM disk and we could not determine the cause for these failures. Thus, the time and RAM usage for grlBWT was obtained without using the RAM disk. BigBWT and eGap are external algorithms as well, but their implementations lack an option to change the directory of temporary files; thus, we obtain their times and RAM usages without using the RAM disk.

Our implementation of IBB is able to perform only one of the two levels of the last predecessor symbol, so more values for $k$ are possible. We refer to these by $x.5$ values for the parameter $k$.\footnote{The IBB implementation actually takes $2k$ as input.}
We use $k=2.5$ in all our tests as parameter; in \autoref{fig:k_results}, we show an evaluation for parameter $k$. The fastest time IBB achieves on the dataset zymowastewater is $343.0701$ seconds for $k=5.5$. From $2.5$ to $7.5$ all recorded times only differ by $17$ seconds.

The scatter plots in \autoref{fig:time_and_space} show the time in seconds and RAM usage in MB given the file and the approach. We say an algorithm $A$ is superior to another approach $B$ if $A$ uses less time, which means the point of $A$ is left of the point of $B$, and less RAM, so the point of $A$ is lower than the point of $B$. For example, ropebwt is superior to BigBWT on the file TAIR10, because ropebwt uses less time and less RAM compared to BigBWT, so ropebwts point is in the left of and below the point of BigBWT.

Our test shows that IBB is the fastest construction algorithm on all datasets. On the influenza dataset, which has longer but fewer sequences, ropebwt2 and IBB both take 61 to 62 seconds, with IBB being slightly faster. On the other datasets, IBB is between $11.33\%$ (HGChr14) and $36.16\%$ (zymowastewater) faster than the second fastest algorithm. The second fastest algorithms are divsufsort on zymowastewater, ropebwt3 on SRR11092057, and ropebwt2 on the remaining datasets. Using the time-optimal parameter $k=5.5$ from \autoref{fig:k_results}, IBB is $40.85\%$ faster than divsufsort on zymowastewater.  

Ropebwt uses the lowest amount of RAM on all datasets followed by either ropebwt2 or grlBWT. The latter two mostly offer pareto-optimal space-time tradeoffs for BWT construction. They are always followed by IBB, which is then superior to any other tested approach in both, construction time and RAM usage.

In \autoref{table:external_memory_evaluation} in \autoref{appendix:abbreviations}, we compare the time and external memory used by the external-memory algorithms without using a RAM disk. Of the algorithms in \autoref{tab:bwt_construction_algortihms}, BCR, BigBWT, eGap, grlBWT, r-pfbwt, and IBB are external memory algorithms. We analyzed the total amount of input and output written by the algorithms, obtained as the sum of the io\_in and io\_out value. IBB has a much higher amount of input and output; nevertheless, except for the GRCm39 and JAGHKL01 datasets, IBB is the fastest external memory approach.

\section{Conclusion}

We presented IBB, a BWT construction algorithm designed for length-diverse DNA datasets. IBB employs right alignment, tree-based data structures, fine buckets, and one-step radix sorting. Our experiments show that IBB outperforms existing state-of-the-art BWT construction algorithms by being 10\% to 40\% faster on most datasets while maintaining competitive memory consumption.

\appendix
\newpage

%
%

\bibliography{biblio}

\section{List of Symbols} \label{appendix:abbreviations}

\begin{longtable}{lp{10cm}}
    Symbol & Explanation \\\hline
\endhead
    $S$ & A single string. \\
    $\Sigma$ & The alphabet for strings. \\
    $rank_S(x, c)$ & Number of occurrences of $c$ in $S[0,x-1]$. \\
    $count_S(c)$ & Number of characters smaller than $c$ in $S$. \\
    $BWT(W)$ & Burrows-Wheeler transform of $W$. \\
    $(S_i)_{0 \leq i < m}$ & Collection of $m$ words. \\
    $j, z$ & Indices used for words. \\
    $M$ & Maximal length of words in the collection of words. \\
    $W$ & String with $\$_i$ to define the BWT of a collection. \\
    $t$ & Current iteration (counter). \\
    $LF(i)$ & Last-to-First mapping of position $i$. Used to reverse the Burrows-Wheeler transform. \\
    $bwt(t)$ & Partial $BWT$ after iteration $t$. \\
    $(W_i)_{0 \leq i < m}$ & Collection of words obtained from $(S_i)_{0 \leq i < m}$ by reversing and right aligning. \\
    $P_j(t)$ & Insert position of word $W_j$ in iteration $t$. We sometimes call this insert position global to make the difference to the other insert positions more verbose.\\
    $\alpha(t)$ & Number of active words at iteration $t$. \\
    $SB(t)$ & Bitvector at iteration $t$. $SB(t)[i]=1$ if and only if word $W_i$ is active. \\
    $B_c$ & Level-1 bucket for predecessor symbol $c$. \\
    $k$ & Parameter to IBB. Defines the depth of the trees as $2k-2$ and the length of the predecessor sequence. \\
    $B_{c_{1} \dots c_{k}}$ & Level-k bucket for predecessor sequence $c_{1} \dots c_{k}$ \\
    $AW_j$ & Constructed word that includes the prefixed $A^k$. $AW_j = (A^k \cdot W_{j})$ \\
    $D$ & Any predecessor sequence. Typically of length $k$, yet not exclusively. \\
    $C_W(D)$ & Predecessor sequence counting based on $AW_j$ and length $k$. \\
    $D_j(t)$ & Predecessor sequence of word $W_j$ in iteration $t$. \\
    $P_j^{c}(t)$ & Insert position of word $W_j$ in iteration $t$ relative to the bucket $B_c$ for a single character $c$. \\
    $P_j^{D_j(t)}(t)$ & Insert position of word $W_j$ in iteration $t$ relative to the bucket $B_{D_j(t)}$ for predecessor sequence $D_j(t)$. \\
    $T$ & Array to save the tree. \\
    $rank1(j, t)$ & Rank of inserted character $W_j[t]$ at position $P_{j}^{W_j[t-1]}(t)$ on the string $B_{W_j[t-1]}$, which is the level-1 bucket, so the whole string within the $W_j[t-1]$ tree. \\
    $R_D.c$ & Accumulators during tree processing. $D$ is the predecessor sequence (not necessarily of length $k$), $c \in \Sigma$. \\
    $rank\text{k}(j, t)$ & Rank of inserted character $W_j[t]$ at position $P_{j}^{{D_j(t)}}(t)$ on the string $B_{D_j(t)}$, which is the level-k bucket. \\
    $F_i(B_D)$ & File with index $i$ associated to the bucket $B_D$. \\
    $L_D(t)$ & Index of file to write of bucket $B_D$ in iteration $t$. \\
    $i, x$ & Indices with minimal scope.
\end{longtable}

\section{Evaluation Results}\label{appendix:test_results}

\newcommand{\tabwidth}{0.48\linewidth}

\begin{table}[!h]
    \caption{The following tables list the results for the tests using RAM disk.}
\begin{subtable}{\tabwidth}
\caption{TAIR10}
\begin{tabular}{lrr}
	Approach & Time (s) & max-rss (MB) \\\hline
	BCR & – & – \\
	BigBWT & 53.1195 & 1,357.90 \\
	divsufsort & 10.4938 & 1,049.59 \\
	eGap & 16.0234 & 4,078.12 \\
	grlBWT & 24.9048 & 669.00 \\
	gsufsort & 17.2666 & 1,223.87 \\
	IBB & 7.1295 & 383.66 \\
	r-pfbwt & 87.1017 & 2,401.18 \\
	ropebwt & 36.9868 & 71.68 \\
	ropebwt2 & 10.3008 & 350.90 \\
	ropebwt3 & 12.1775 & 1,570.97 \\
\end{tabular}
\end{subtable}
\begin{subtable}{\tabwidth}
\caption{GRCm39}
\begin{tabular}{lrr}
	Approach & Time (s) & max-rss (MB) \\\hline
	BCR & – & – \\
	BigBWT & 1,282.7032 & 52,893.75 \\
	divsufsort & 305.4121 & 21,796.86 \\
	eGap & – & – \\
	grlBWT & 703.2101 & 4,501.62 \\
	gsufsort & 430.7094 & 26,563.23 \\
	IBB & 167.4545 & 6,685.82 \\
	r-pfbwt & 2,499.6059 & 51,853.26 \\
	ropebwt & 1,712.1969 & 2,569.87 \\
	ropebwt2 & 223.1947 & 6,585.71 \\
	ropebwt3 & 227.3670 & 45,098.62 \\
\end{tabular}
\end{subtable}
\end{table}

\begin{table}[!h]
\begin{subtable}{\tabwidth}
\setcounter{subtable}{2}
\caption{GRCh38}
\begin{tabular}{lrr}
	Approach & Time (s) & max-rss (MB) \\\hline
    BCR & – & – \\
	BigBWT & 1,571.8187 & 62,954.50 \\
	divsufsort & 354.5520 & 26,577.39 \\
	eGap & – & – \\
	grlBWT & 745.3975 & 4,857.22 \\
	gsufsort & 498.4121 & 26,630.89 \\
	IBB & 207.5513 & 8,989.36 \\
	r-pfbwt & 2,981.6389 & 62,886.01 \\
	ropebwt & 1,882.8899 & 2,938.64 \\
	ropebwt2 & 256.6377 & 8,056.41 \\
	ropebwt3 & 258.3919 & 51,735.07 \\
\end{tabular}
\end{subtable}
\begin{subtable}{\tabwidth}
\caption{JAGHKL01}
\begin{tabular}{lrr}
	Approach & Time (s) & max-rss (MB) \\\hline
	BCR & – & – \\
	BigBWT & – & – \\
	divsufsort & – & – \\
	eGap & – & – \\
	grlBWT & 2,091.7559 & 9,722.14 \\ 
	gsufsort & – & – \\
	IBB & 930.3123 & 29,011.98 \\
	r-pfbwt & – & – \\
	ropebwt & 7,177.5101 & 5,251.80 \\
	ropebwt2 & 1,046.8844 & 17,736.01 \\
	ropebwt3 & – & – \\
\end{tabular}
\end{subtable}
\end{table}

\begin{table}[!h]
\begin{subtable}{\tabwidth}
\setcounter{subtable}{4}
\caption{SRR11092057}
\begin{tabular}{lrr}
	Approach & Time (s) & max-rss (MB) \\\hline
	BCR & – & – \\
	BigBWT & 356.1337 & 8,709.25 \\
	divsufsort & 80.6930 & 6,741.88 \\
	eGap & 107.6240 & 7,862.65 \\
	grlBWT & 158.0011 & 1,790.23 \\
	gsufsort & 129.9375 & 8,169.07 \\
	IBB & 51.5733 & 2,176.63 \\
	r-pfbwt & 808.8667 & 14,901.21 \\
	ropebwt & 485.2979 & 691.24 \\
	ropebwt2 & 203.0154 & 1,588.63 \\
	ropebwt3 & 71.3488 & 8,364.67 \\
\end{tabular}
\end{subtable}
\begin{subtable}{\tabwidth}
\caption{influenza}
\begin{tabular}{lrr}
	Approach & Time (s) & max-rss (MB) \\\hline
	BCR & – & – \\
	BigBWT & 138.3262 & 4,552.87 \\
	divsufsort & 121.6796 & 9,660.54 \\
	eGap & 134.2194 & 11,183.71 \\
	grlBWT & 121.6891 & 783.07 \\
	gsufsort & 170.1249 & 11,595.05 \\
	IBB & 61.4567 & 3,747.00 \\
	r-pfbwt & 83.3559 & 3,797.16 \\
	ropebwt & 456.7642 & 92.92 \\
	ropebwt2 & 61.6558 & 1,294.27 \\
	ropebwt3 & 70.5649 & 11,686.07 \\
\end{tabular}
\end{subtable}
\end{table}

\begin{table}[!ht]
\begin{subtable}{\tabwidth}
\setcounter{subtable}{6}
\caption{HGChr14}
\begin{tabular}{lrr}
	Approach & Time (s) & max-rss (MB) \\\hline
	BCR & – & – \\
	BigBWT & 453.5586 & 11,790.17 \\
	divsufsort & 112.1837 & 8,980.93 \\
	eGap & 142.3352 & 10,236.23 \\
	grlBWT & 167.9500 & 1,341.46 \\
	gsufsort & 171.9143 & 11,349.88 \\
	IBB & 77.8972 & 6,255.21 \\
	r-pfbwt & 1,035.8426 & 18,928.85 \\
	ropebwt & 568.6316 & 495.94 \\
	ropebwt2 & 276.8750 & 1,673.65 \\
	ropebwt3 & 87.8484 & 10,737.50 \\
\end{tabular}
\end{subtable}
\begin{subtable}{\tabwidth}
\caption{zymowastewater}
\begin{tabular}{lrr}
	Approach & Time (s) & max-rss (MB) \\\hline
	BCR & – & – \\
	BigBWT & – & – \\
	divsufsort & 763.8530 & 56,533.53 \\
	eGap & 929.7517 & 56,763.04 \\
	grlBWT & – & – \\
	gsufsort & 1,103.9456 & 59,621.83 \\
	IBB & 370.2614 & 21,948.47 \\
	r-pfbwt & – & – \\
	ropebwt & 4,788.0798 & 5,051.52 \\
	ropebwt2 & 2,049.8674 & 11,528.41 \\
	ropebwt3 & – & – \\
\end{tabular}
\end{subtable}
\end{table}

\begin{table}[!ht]
    \setcounter{table}{3}
    \caption{The following tables list the results for external memory approaches without using a RAM disk.}\label{table:external_memory_evaluation}
\begin{subtable}{\tabwidth}
    \caption{TAIR10}
    \begin{tabular}{lrr}
        Approach & Time (s) & IO (MB) \\\hline
        BCR & – & – \\
        BigBWT & 54.7431 & 255.72 \\
        eGap & 16.7038 & 0.17 \\
        grlBWT & 24.9048 & 1,355.56 \\
        IBB & 14.7806 & 15,064.05 \\
        r-pfbwt & 92.4510 & 338.87 \\
    \end{tabular}
    \end{subtable}
    \begin{subtable}{\tabwidth}
    \caption{GRCm39}
    \begin{tabular}{lrr}
        Approach & Time (s) & IO (MB) \\\hline
        BCR & – & – \\
        BigBWT & 1,371.5356 & 6,586.37 \\
        eGap & 447.5178 & 2,130.45 \\
        grlBWT & 703.2101 & 39,075.34 \\
        IBB & 449.2580 & 511,059.49 \\
        r-pfbwt & 2,714.2478 & 12,496.36 \\
    \end{tabular}
    \end{subtable}
\end{table}

\begin{table}[!ht]
    \begin{subtable}{\tabwidth}
    \setcounter{subtable}{2}
    \caption{GRCh38}
    \begin{tabular}{lrr}
        Approach & Time (s) & IO (MB) \\\hline
        BCR & – & – \\
        BigBWT & 1,572.0601 & 7,696.68 \\
        eGap & 570.0827 & 2,952.55 \\
        grlBWT & 745.3975 & 60,962.59 \\
        IBB & 493.8601 & 557,734.03 \\
        r-pfbwt & 3,275.8786 & 20,418.35 \\
    \end{tabular}
    \end{subtable}
    \begin{subtable}{\tabwidth}
    \caption{JAGHKL01}
    \begin{tabular}{lrr}
        Approach & Time (s) & IO (MB) \\\hline
        BCR & – & – \\
        BigBWT & – & – \\
        eGap & – & – \\
        grlBWT & 2,091.7559 & 185,232.58 \\
        IBB & 2,512.7515 & 4,235,386.38 \\
        r-pfbwt & – & – \\
    \end{tabular}
    \end{subtable}
\end{table}

\begin{table}[!ht]
\begin{subtable}{\tabwidth}
    \setcounter{subtable}{4}
\caption{SRR11092057}
\begin{tabular}{lrr}
	Approach & Time (s) & IO (MB) \\\hline
	BCR & – & – \\
	BigBWT & 355.0525 & 1,952.04 \\
	eGap & 109.0181 & 624.06 \\
	grlBWT & 158.0011 & 7,629.49 \\
	IBB & 66.8494 & 34,861.29 \\
	r-pfbwt & 847.7592 & 3,129.21 \\
\end{tabular}
\end{subtable}
\begin{subtable}{\tabwidth}
\caption{influenza}
\begin{tabular}{lrr}
	Approach & Time (s) & IO (MB) \\\hline
	BCR & – & – \\
	BigBWT & 148.4186 & 1,834.74 \\
	eGap & 139.4808 & 970.80 \\
	grlBWT & 121.3235 & 3,362.62 \\
	IBB & 66.0257 & 23,235.80 \\
	r-pfbwt & 95.4392 & 445.43 \\
\end{tabular}
\end{subtable}
\end{table}

\begin{table}[!ht]
\begin{subtable}{\tabwidth}
    \setcounter{subtable}{6}
\caption{HGChr14}
\begin{tabular}{lrr}
	Approach & Time (s) & IO (MB) \\\hline
	BCR & – & – \\
	BigBWT & 462.9071 & 2,586.88 \\
	eGap & 147.4293 & 402.06 \\
	grlBWT & 167.9500 & 10,243.43 \\
	IBB & 91.0889 & 45,861.17 \\
	r-pfbwt & 1,119.1047 & 3,431.50 \\
\end{tabular}
\end{subtable}
\begin{subtable}{\tabwidth}
\caption{zymowastewater}
\begin{tabular}{lrr}
	Approach & Time (s) & IO (MB) \\\hline
	BCR & – & – \\
	BigBWT & – & – \\
	eGap & 933.8026 & 5,264.78 \\
	grlBWT & 1,475.5007 & 92,775.28 \\
	IBB & 619.8390 & 443,232.04 \\
	r-pfbwt & – & – \\
\end{tabular}
\end{subtable}
\end{table}

\end{document}